\documentclass[aps,prd,twocolumn,amsmath,amssymb,showpacs]{revtex4}

\usepackage{graphicx}
\usepackage{dcolumn}
\usepackage{bm}

\begin{document}

\title{Regge poles of the Schwarzschild black hole: a WKB approach}

\author{Yves D\'ecanini}
\email{decanini@univ-corse.fr}
\author{Antoine Folacci}
\email{folacci@univ-corse.fr} \affiliation{
UMR CNRS 6134 SPE, Equipe Physique Th\'eorique, \\
Universit\'e de Corse, Facult\'e des Sciences, BP 52, 20250 Corte,
France}

\date{\today}

\begin{abstract}
We provide simple and accurate analytical expressions for the Regge
poles of the Schwarzschild black hole. This is achieved by using
third-order WKB approximations to solve the radial wave equations
for spins 0, 1, and 2. These results permit us to obtain
analytically the dispersion relation and the damping of the
``surface waves" lying close to the photon sphere of the
Schwarzschild black hole and which generate the weakly damped
quasinormal modes of its spectrum. Our results could be helpful in
order to simplify considerably the description of wave scattering
from the Schwarzschild black hole as well as the analysis of the
gravitational radiation created in many black hole processes.
Furthermore, the existence of nonlinear dispersion relations for the
photons propagating close to the photon sphere could have also
important consequences in the context of gravitational lensing.
\end{abstract}


\maketitle

\section{Introduction}

Since the sixties, mainly under the impetus of Nussenzveig in
electromagnetism and Regge in quantum physics,
asymptotic/semiclassical techniques which use analytic continuation
of partial-wave representations have been developed to understand
scattering theory and the associated resonance phenomenons. Together
these techniques form the complex angular momentum (CAM) method. The
CAM method originates from the pioneering work of Watson
\cite{Watson18} dealing with the propagation and diffraction of
radio waves around the earth (see also the modification proposed by
Sommerfeld \cite{Sommerfeld49} of the Watson approach). Today, the
CAM method has been successfully introduced in various domains of
physics. For reviews of the CAM method we refer to the monographs of
Newton \cite{New82}, Nussenzveig \cite{Nus92}, Grandy \cite{Grandy},
Collins \cite{Collins77} and Gribov \cite{Gribov69} as well as to
references therein for various applications in quantum mechanics,
nuclear physics, particle and high-energy physics, electromagnetism,
optics, and seismology.

The CAM method is based on the dual structure of the $S$-matrix
associated with a given scattering problem. Indeed, the $S$-matrix
is a function of both the frequency $\omega$ and the angular
momentum index $\ell$. It can be analytically extended into the
complex $\omega$ plane as well as into the complex $\ell$ plane (CAM
plane). Its poles lying in the fourth quadrant of the complex
$\omega$ plane are the complex frequencies of the resonant modes. In
other words, the behavior of the $S$-matrix in the complex $\omega$
plane permits us to investigate resonance phenomenons. The structure
of the $S$-matrix in the complex $\ell$ plane allows us -- by using
integration contour deformations, Cauchy's Theorem, and asymptotic
analysis -- to provide a semiclassical description of scattering in
terms of geometric and diffractive rays. In that context, the poles
of the $S$-matrix lying in the first quadrant of the CAM plane (the
so-called Regge poles) are associated with diffraction. Of course,
when a connection between these two faces of the $S$-matrix can be
established, resonance aspects of scattering are then
semiclassically interpreted.

More precisely, for spherically symmetric scattering problems, the
connection between the resonant approach of scattering and its Regge
pole interpretation is a consequence of the following
considerations:

\quad i) For a given value $\ell \in {\bf N}$ of the angular
momentum index, the resonances, i.e., the complex poles of the
matrix element $S_\ell (\omega)$, are of the form $\omega_{\ell
n}=\omega^{(o)}_{\ell n}- i\Gamma_{\ell n}/2$, where $n\in {\bf N}$
and, in the immediate neighborhood of $\omega_{\ell p}$,
$S_\ell(\omega)$ has the Breit-Wigner form, i.e.,
\begin{equation}\label{BW}
S_\ell(\omega ) \propto \frac{\Gamma _{\ell n}/2}{\omega
-\omega^{(o)}_{\ell n}+i\Gamma _{\ell n}/2}.
\end{equation}

\quad ii) For a given value $\omega >0$ of the frequency, the Regge
poles, i.e., the complex poles of the matrix element $S_{\lambda
-1/2}(\omega)$, can be expressed as function of the form $\lambda_n
(\omega)$. The curves traced out in the CAM plane by the Regge poles
as a function of the frequency $\omega$ are the so-called Regge
trajectories. They permit us to interpret Regge poles in terms of
diffractive rays or ``surface waves": $\mathrm{Re} \, \lambda_n
(\omega)$ provides the dispersion relation for the $n$th ``surface
wave", while $\mathrm{Im} \, \lambda_n (\omega)$ corresponds to its
damping.

\quad (iii) From the Regge trajectories, we can semiclassically
construct the resonance spectrum. There exists a first semiclassical
formula (a Bohr-Sommerfeld-type quantization condition) which
provides the location of the excitation frequencies
$\omega^{(0)}_{\ell n}$ of the resonances generated by $n$th
``surface wave":
\begin{equation}\label{sc1}
\mathrm{Re}  \, \lambda_n \left(\omega^{(0)}_{\ell n} \right)= \ell
+ 1/2,  \qquad \ell \in {\bf N}.
\end{equation}
There exists a second semiclassical formula which provides the
widths of these resonances
\begin{equation}\label{sc2} \frac{\Gamma _{\ell n}}{2}= \left.  \frac{\mathrm{Im} \,
\lambda_n (\omega )[d/d\omega \, \mathrm{Re}\, \lambda_n (\omega )
]}{[d/d\omega \, \mathrm{Re} \, \lambda_n (\omega ) ]^2 + [d/d\omega
\, \mathrm{Im} \, \lambda_n (\omega ) ]^2 } \right|_{\omega
=\omega^{(0)}_{\ell n}}
\end{equation}
and which reduces, in the frequency range where the condition $|
d/d\omega \,\mathrm{Re} \, \lambda_n (\omega )  | \gg |d/d\omega \,
\mathrm{Im}\, \lambda_n (\omega )  |$ is satisfied, to
\begin{equation}\label{sc4}
\frac{\Gamma _{\ell n}}{2}= \left.  \frac{\mathrm{Im} \, \lambda_n
(\omega )}{d/d\omega \, \ \mathrm{Re} \, \lambda_n (\omega )  }
\right|_{\omega =\omega^{(0)}_{\ell n}}.
\end{equation}

Recently, by using CAM techniques or in other words the Regge pole
machinery, we have established that the weakly damped quasinormal
mode (QNM) complex frequencies of the Schwarzschild black hole of
mass $M$ are Breit-Wigner-type resonances generated by a family of
``surface waves" lying on its photon sphere at $r=3M$
\cite{DecaniniFJ_cam_bh}. More precisely, by noting that each
``surface wave" is associated with a Regge pole of the $S$-matrix of
the Schwarzschild black hole
\cite{Andersson1,Andersson2,DecaniniFJ_cam_bh}, we have been able to
numerically construct the spectrum of the QNM complex frequencies
from the associated Regge trajectories. In fact, in this way, from
CAM methods, we have given meaning to an appealing and physically
intuitive interpretation of the Schwarzschild black hole QNMs
suggested by Goebel in 1972 \cite{Goebel}, i.e., that they could be
interpreted in terms of gravitational waves in spiral orbits close
to the unstable circular photon orbit at $r=3M$ which decay by
radiating away energy. It should be noted that alternative
implementations of the Goebel interpretation have been developed
(see Refs.~\cite{FerrariMashhoon84,Mashhoon85,Stewart89,
AnderssonOnozawa96,ZerbiniVanzo2004,CardosoMirandaBertietal2009})
but they are mainly based on the concepts of geodesic and bundle of
geometrical rays while our CAM analysis is based on wave physics.

In Sec.~2 of this short note, we provide analytical expressions for
the Regge poles corresponding to the scalar field (spin 0), the
electromagnetic field (spin 1) and the gravitational perturbations
(spin 2) propagating on the Schwarzschild background. This is
achieved by using the WKB approach developed in the context of the
determination of the QNMs by Schutz and Will \cite{SchutzWill} and
by Will and Iyer \cite{Iyer1,Iyer2} (see also
Ref.~\cite{BenderOrszag1978} for general aspects of WKB theory and
for particular aspects connected with eigenvalue problems of the
type considered here). Our WKB results permit us to describe very
accurately the Regge trajectories of the Schwarzschild black hole
and to recover from the semiclassical formulas (\ref{sc1}) and
(\ref{sc4}) well-known analytical expressions for the QNM complex
frequencies. It should be noted that in
Ref.~\cite{DecaniniFJ_cam_bh}, from numerical considerations, we
have provided a coarse formula for the Regge poles of the
Schwarzschild black hole. It has been sufficient in order to obtain
a ``surface wave" interpretation of the QNM complex frequencies but
it does not take into account the spin dependent aspects of the
problem as well as the fact that (i) the dispersion relation of the
$n$th ``surface wave" is nonlinear and depends on the index $n$ and
that (ii) the damping of the $n$th ``surface wave" is frequency
dependent. In the present paper, we have derived analytical
expressions for the Regge poles of the Schwarzschild black hole from
mathematical considerations. The results displayed in
Ref.~\cite{DecaniniFJ_cam_bh} are the leading-order counterpart for
very high frequencies of our new results which are much more
accurate and which now permit us to describe precisely the behavior
of the Regge poles and of the associated ``surface waves" in a very
large range of frequencies. In Sec.~3, we briefly conclude by
considering some possible consequences and applications of our
results.

\section{Regge poles, Regge trajectories, and WKB results}

The wave equations for the scalar field, for the electromagnetic
field, and for the gravitational perturbations propagating on the
Schwarzschild black hole reduce, after separation of variables, to
the Regge-Wheeler equation (a one-dimensional Schr\"odinger
equation)
\begin{equation}\label{RW}
\frac{d^2 \Phi_\ell}{dr_*^2} + \left[ \omega^2 - V_\ell(r)\right]
\Phi_\ell=0
\end{equation}
where the effective potential $V_\ell(r)$ is given by
\begin{equation}\label{EffectivePot_spin_s}
V_\ell(r) = \left(\frac{r-2M}{r} \right) \left[
\frac{\ell(\ell+1)}{r^2} +\frac{2(1-s^2)M}{r^3}\right].
\end{equation}
Here, $s$ denotes the spin of the field considered, $\ell \in {\bf
N}$ is the ordinary angular momentum index, while $r$ and
$r_*=r+2M\ln \left(r/2M - 1 \right) + \mathrm{const}$ are,
respectively, the standard radial Schwarzschild coordinate and the
Regge-Wheeler tortoise coordinate, and we have furthermore assumed a
harmonic time dependence $e^{-i\omega t}$.

For a given angular momentum index $\ell$, the $S$-matrix element
$S_\ell (\omega)$ is obtained by seeking the solutions of the
Regge-Wheeler equation (\ref{RW}) which have a purely ingoing
behavior at the event horizon $r=2M$, i.e. which satisfy
\begin{equation}\label{bc1}
\Phi_\ell (r) \underset{r_* \to -\infty}{\sim}
e^{-i\omega r_* }
\end{equation}
and which, at spatial infinity $r \to +\infty$, have an asymptotic
behavior of the form
\begin{equation}\label{bc2}
\Phi_\ell(r) \underset{r_* \to +\infty}{\sim}
\frac{1}{T_\ell(\omega)}e^{-i\omega r_* + i\ell\pi/2}-
\frac{S_\ell(\omega)}{T_\ell(\omega)} e^{+i\omega r_* - i\ell\pi/2}.
\end{equation}

\begin{figure}
\includegraphics[height=6.0cm,width=8cm]{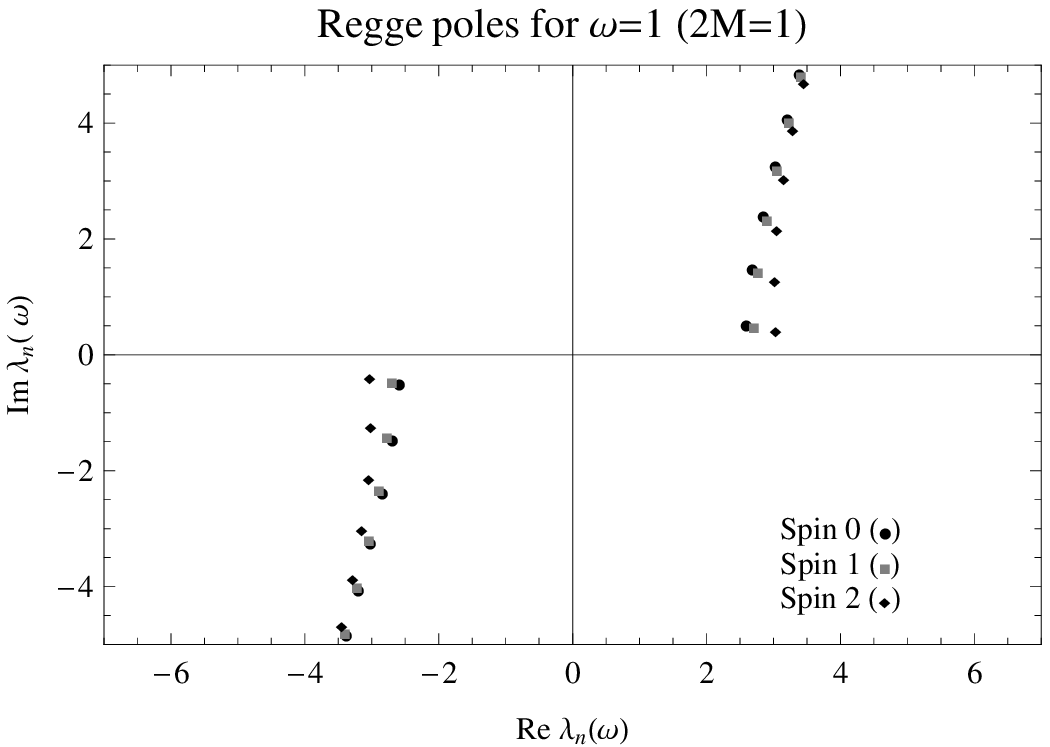}
\includegraphics[height=6.0cm,width=8cm]{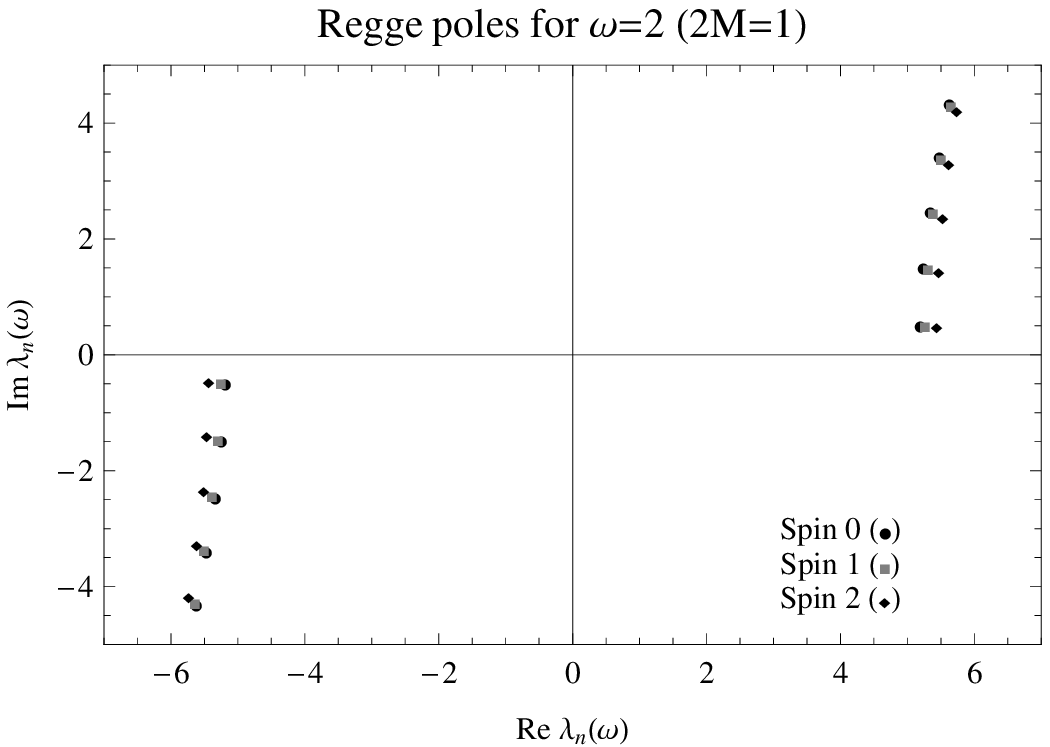}
\caption{\label{fig:PRspins012} Regge pole distributions for
$\omega=1$ and $\omega=2$ ($2M=1$). Scalar and electromagnetic
fields as well as gravitational perturbations are considered.}
\end{figure}

For $\omega >0$, the Regge poles of the $S$-matrix are the complex
values $\lambda$ for which, both $S_{\lambda -1/2}(\omega)$ and
$T_{\lambda -1/2}(\omega)$ have a simple pole but $S_{\lambda
-1/2}(\omega)/T_{\lambda -1/2}(\omega)$ is regular
\cite{DecaniniFJ_cam_bh}. They constitute a family
$\lambda_n(\omega)$ with the index $n=1,2, \dots $ permitting us to
distinguish between the different poles. The corresponding modes are
the solutions of the radial wave equation (\ref{RW}) which are
purely outgoing at infinity and purely ingoing at the horizon. They
are the ``Regge modes" of the Schwarzschild black hole. It should be
noted that the QNMs of the Schwarzschild black hole and the
associated complex frequencies are defined in an analogous way with
moreover a fundamental difference. Indeed, they are obtained for
fixed values of $\ell \in {\bf N}$ and exist only for complex values
of $\omega$.

\begin{figure}
\includegraphics[height=6cm,width=8cm]{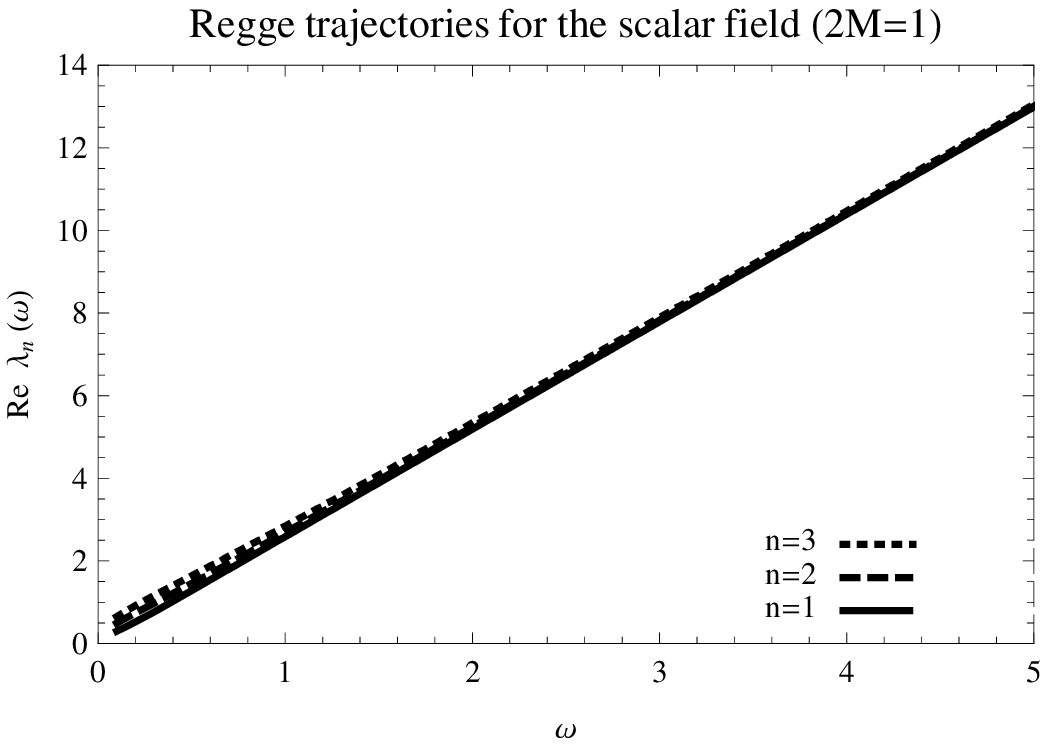}
\includegraphics[height=6cm,width=8cm]{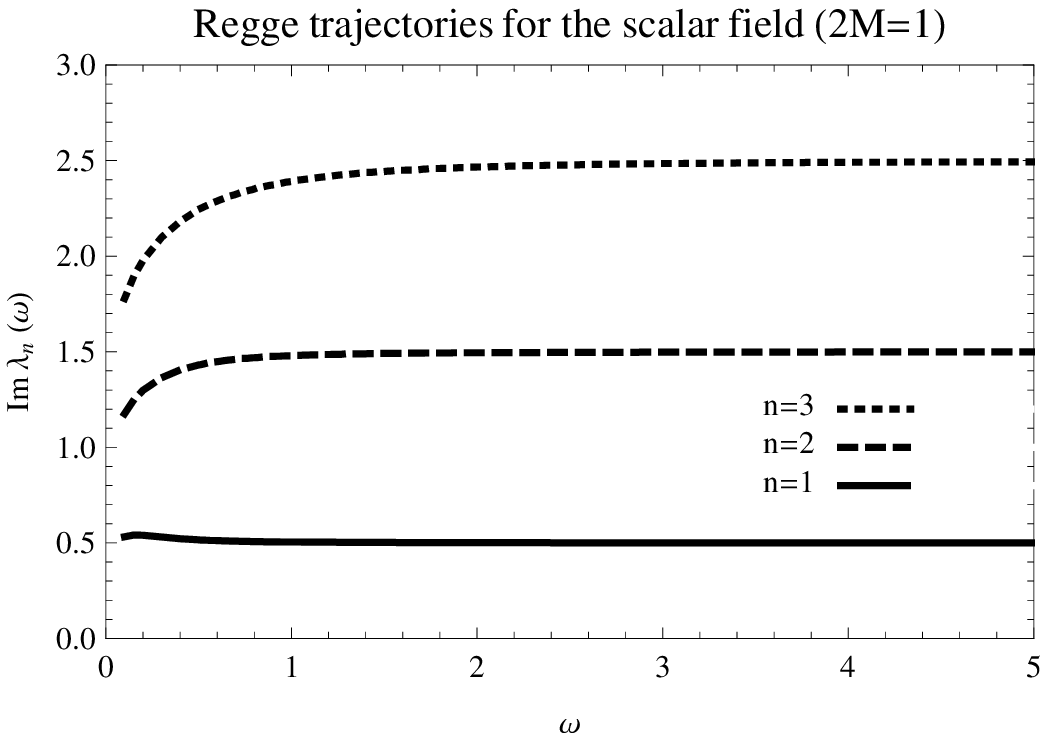}
 \caption{\label{fig:TRspin0}Regge trajectories for the scalar field.
 The Regge poles $\lambda_n(\omega)$ ($n=1,2,3$) are followed
for $\omega=0\to 5$ ($2M=1$). }
\end{figure}

\begin{figure}
\includegraphics[height=6cm,width=8cm]{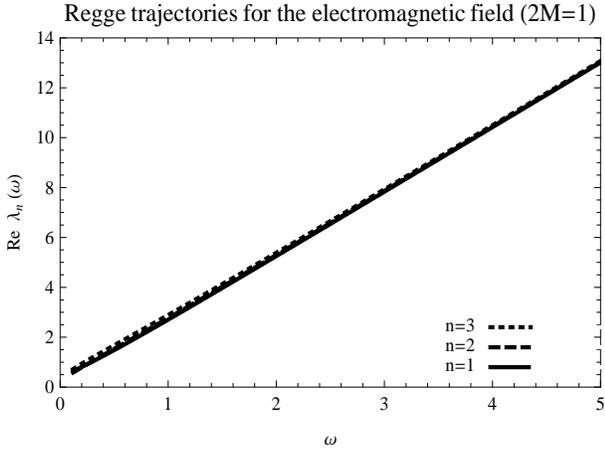}
\includegraphics[height=6cm,width=8cm]{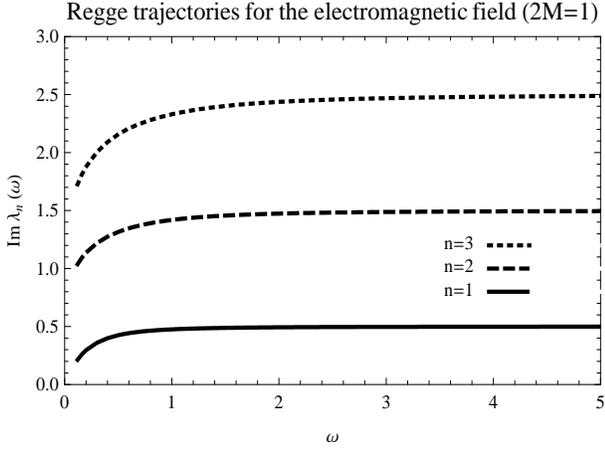}
 \caption{\label{fig:TRspin1}Regge trajectories for the electromagnetic field.
 The Regge poles $\lambda_n(\omega)$ ($n=1,2,3$) are followed
for $\omega=0\to 5$ ($2M=1$). }
\end{figure}

\begin{figure}
\includegraphics[height=6cm,width=8cm]{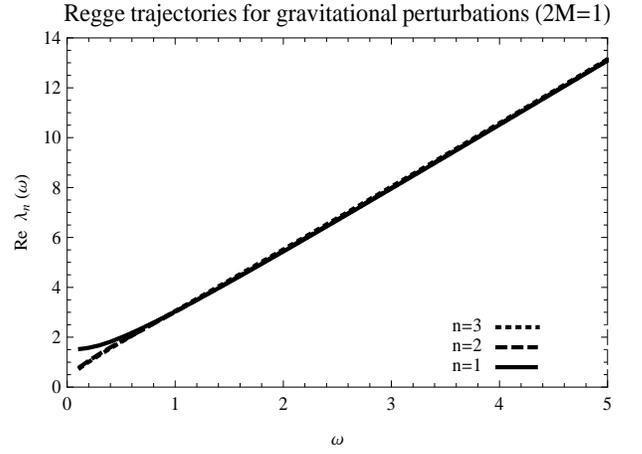}
\includegraphics[height=6cm,width=8cm]{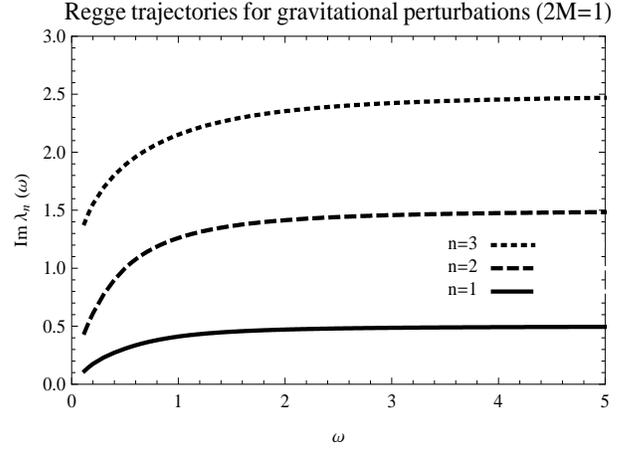}
 \caption{\label{fig:TRspin2}Regge trajectories for gravitational perturbations.
 The Regge poles $\lambda_n(\omega)$ ($n=1,2,3$) are followed
for $\omega=0\to 5$ ($2M=1$). }
\end{figure}

In Ref.~\cite{DecaniniFJ_cam_bh}, we have noted that the Regge poles
can be determined by using, {\it mutatis mutandis}, the method
developed by Leaver \cite{LeaverI} for the search of the QNM complex
frequencies and we have furthermore numerically implemented this
method from the Hill determinant approach of Majumdar and
Panchapakesan \cite{mp}. For the Regge poles lying close to the real
axis of the CAM plane, the ``exact" results we can obtain
numerically are accurate ones. Figure \ref{fig:PRspins012} exhibits
the distribution of Regge poles for two different values of the
frequency $\omega$. On this figure, the splitting of the Regge poles
with the spin $s$ and the index $n$ as well as the migration of the
first Regge poles corresponding to spins 0, 1 and 2 can be observed.
This migration can be more precisely described by studying the Regge
trajectories, i.e., the curves traced out in the CAM plane by the
functions $\lambda_n(\omega)$ for $\omega \in [0, +\infty[$. In
Figs.~\ref{fig:TRspin0}-\ref{fig:TRspin2} we have displayed the
Regge trajectories for the first three Regge poles associated with
the scalar field, the electromagnetic field and the gravitational
perturbations. As we have already noted in
Ref.~\cite{DecaniniFJ_cam_bh}, $\mathrm{Re} \, \lambda_n(\omega)$
and $\mathrm{Im} \, \lambda_n(\omega)$ respectively denote the
azimuthal propagation constant (i.e., the dispersion relation) and
the damping constant of the $n$th ``surface wave" lying on the
photon sphere. So Figs.~\ref{fig:TRspin0}-\ref{fig:TRspin2} permit
us to note that, at first sight, (i) the global behavior of these
``surface waves" is rather independent of the spin considered (such
result is only valid for high frequencies) and (ii) for a given
spin, the dispersion relation seems to be independent of the
``surface wave" considered  (such result is also only valid for high
frequencies) and the ``surface wave" corresponding to the first
Regge pole has the weakest attenuation and therefore must play the
most important role in scattering and in radiative processes and
must contribute significantly to the resonance mechanism.


The WKB approach developed in the context of the determination of
the QNMs by Schutz and Will \cite{SchutzWill} and by Will and Iyer
\cite{Iyer1,Iyer2} (see also Ref.~\cite{BenderOrszag1978}) can be
easily adapted in order to obtain analytical approximations for the
Regge poles. Indeed, as we have already noted, Regge modes and QNMs
both satisfy the same wave equation and the same boundary conditions
but with different interpretations for the angular momentum index
and the frequency. By using third-order WKB approximations
\cite{Iyer1,Iyer2} for the Regge modes, we can find that the Regge
poles $\lambda$ are the complex solutions of the equation
\begin{eqnarray}\label{WKB_RP_1}
& &\omega^2=\left[V_0(\lambda) + {[-2V_0^{(2)}(\lambda)]}^{1/2} \,
{\overline \Lambda}(\lambda,n) \right] \nonumber \\ && \qquad - i \,
\alpha (n) \, {[-2V_0^{(2)}(\lambda)]}^{1/2}[1+ {\overline
\Omega}(\lambda ,n)]
\end{eqnarray}
with $\omega >0$ and $n=1,2,3, \dots $. Here
\begin{subequations}\label{WKB_RP_2}
\begin{eqnarray}\label{WKB_RP_2a}
& & {\overline
\Lambda}(\lambda,n)=\frac{1}{{[-2V_0^{(2)}(\lambda)]}^{1/2}}
\left[\frac{1}{8}
\frac{V_0^{(4)}(\lambda)}{V_0^{(2)}(\lambda)}\left( \frac{1}{4}+
\alpha(n)^2 \right) \right. \nonumber \\
& & \qquad \left.  -\frac{1}{288}
\left(\frac{V_0^{(3)}(\lambda)}{V_0^{(2)}(\lambda)}\right)^2\left(
7+ 60 \, \alpha(n)^2 \right) \right]
\end{eqnarray}
and
\begin{eqnarray}\label{WKB_RP_2b}
& &{\overline \Omega}(\lambda,n)=\frac{1}{[-2V_0^{(2)}(\lambda)]} \times  \nonumber \\
&& \qquad \left[\frac{5}{6912}
\left(\frac{V_0^{(3)}(\lambda)}{V_0^{(2)}(\lambda)}\right)^4\left(
77+ 188 \, \alpha(n)^2 \right) \right. \nonumber \\
&& \qquad \left. -\frac{1}{384}
\left(\frac{{[V_0^{(3)}(\lambda)]}^2V_0^{(4)}(\lambda)}{{[V_0^{(2)}(\lambda)]}^3}\right)
\left(
51+ 100 \, \alpha(n)^2 \right)\right. \nonumber \\
&& \qquad \left. +  \frac{1}{2304}
\left(\frac{V_0^{(4)}(\lambda)}{V_0^{(2)}(\lambda)}\right)^2 \left(
67+ 68 \, \alpha(n)^2 \right) \right. \nonumber \\
&& \qquad \left. +\frac{1}{288} \left(\frac{ V_0^{(3)}(\lambda)
V_0^{(5)}(\lambda)}{{[V_0^{(2)}(\lambda)]}^2}\right) \left( 19+ 28
\, \alpha(n)^2 \right)  \right. \nonumber \\
&& \qquad \left. -\frac{1}{288}
\left(\frac{V_0^{(6)}(\lambda)}{V_0^{(2)}(\lambda)}\right)\left( 5 +
4 \, \alpha(n)^2 \right) \right].
\end{eqnarray}
\end{subequations}
In Eqs.~(\ref{WKB_RP_1}) and (\ref{WKB_RP_2}), we have
introduced the notations
\begin{subequations}
\begin{equation}\label{WKB_RP_3a}
\alpha(n)=n-1/2
\end{equation}
and, for $p \in {\bf N}$,
\begin{equation}\label{WKB_RP_3b}
V_0^{(p)} (\lambda) = \left.  \frac{d^p}{{dr_*}^p }
V_{\lambda-1/2}(r_*)\right|_{r_*={(r_*)}_0}
\end{equation}
\end{subequations}
with ${(r_*)}_0$ which denotes the maximum of the function
$V_{\lambda-1/2}(r_*)$. We can solve Eq.~(\ref{WKB_RP_1}) by
assuming $|\lambda| \gg 1$ as well as $\mathrm{Re} \, \lambda \gg
\mathrm{Im} \, \lambda$. Its solutions constitute a family
$\lambda_n(\omega)$ with $n=1,2,3, \dots $, and we have the
approximation
\begin{eqnarray}\label{PRapproxWKB}
&&\lambda_n(\omega) \approx {[27 (M \omega)^2 +a_n + 2 (n-1/2)^2 \epsilon_n(\omega)]}^{1/2} \nonumber \\
&& \qquad \qquad + i \,(n-1/2) [1+ \epsilon_n(\omega)]
\end{eqnarray}
where
\begin{equation}\label{PRapproxWKB_denote1}
a_n =\frac{2}{3}\left[\frac{5}{12}(n-1/2)^2 + \frac{115}{144} -1
+s^2 \right]
\end{equation}
and
\begin{equation}\label{PRapproxWKB_denote2}
\epsilon_n(\omega) =\frac{b_n}{27 (M \omega)^2 +a_n + (n-1/2)^2}
\end{equation}
with
\begin{equation}\label{PRapproxWKB_denote3}
b_n = -\frac{4}{9}\left[\frac{305}{1728}(n-1/2)^2 +
\frac{5555}{6912} -1 +s^2 \right].
\end{equation}
Equation (\ref{PRapproxWKB}) is the main result of our paper. It
also provides expressions for the dispersion relation and the
damping of the ``surface waves" lying on/close the photosphere of
the Schwarzschild black hole. It is important to note that the
perturbation parameter of the WKB method leading to
Eq.~(\ref{WKB_RP_1}) is not $1/\omega$ with $\omega \to +\infty$
but, in some sense, the distance $\epsilon^{1/2} $ (with formally
$\epsilon \to 0$) between the turning points of Eq.~(\ref{RW})
[i.e., the roots of $Q(r^\ast)= \omega^2 - V_{\ell}(r^\ast)$] and
the location of the peak of $Q(r^\ast)$ (see also
Refs.~\cite{SchutzWill,Iyer1,Iyer2}). As a consequence,
Eq.~(\ref{PRapproxWKB}) is valid in a very large range of frequency
values and not only for $\omega \to +\infty$. Of course, in order to
solve Eq.~(\ref{WKB_RP_1}) and to obtain the expression
(\ref{PRapproxWKB}), we have furthermore assumed that $|\lambda| \gg
1$ as well as $\mathrm{Re} \, \lambda \gg \mathrm{Im} \, \lambda$.
As a consequence, we can expect the accuracy to decrease for very
low frequencies and for higher Regge poles.

\begin{table}
\caption{\label{tab:table1} A sample of Regge pole values for the
scalar field ($2M=1$): Exact versus WKB results.}
\begin{ruledtabular}
\begin{tabular}{cccccc}
&       &\quad Exact  \quad& \quad
Exact \quad&  WKB  &  WKB  \\
 $n $ & $\omega$ &$   \mathrm{Re} \, \lambda_n(\omega)  $
 &  $  \mathrm{Im} \, \lambda_n(\omega)   $
 & $  \mathrm{Re} \, \lambda_n(\omega)    $ &  $ \mathrm{Im} \, \lambda_n(\omega)   $  \\
\hline
1& 0.5 & 1.2828209     & 0.5159935  & 1.2809160    & 0.5180607  \\
 &  &    &   &  (0.15 \% )  &  (  -0.40  \% ) \\
 & 1.0 & 2.5868446    & 0.5047360  & 2.5865154   & 0.5048769 \\
  &  &    &   &    (0.013 \% )  &  (  -0.028 \% ) \\
 & 2.0 & 5.1900754    & 0.5012355  & 5.1900317   & 0.5012441 \\
  &  &    &   &     (0.0008 \% )  &  (  -0.002 \% ) \\
 & 5.0 & 12.9878966    & 0.5002000  & 12.9878938   & 0.5002002 \\
  &  &    &   & (0.00002 \% )  &  ( -0.00004  \% ) \\
 & 8.0 & 20.7830530    & 0.5000782  & 20.7830523   & 0.5000783 \\
  &  &    &   & (0.000003 \% )  &  (  -0.00002  \% ) \\
2& 0.5 & 1.4528181     & 1.4306871  & 1.4448341    & 1.4697683 \\
  &  &    &   &    (0.55 \% )  &  (  -2.73  \% ) \\
 & 1.0 & 2.6883554    & 1.4795865  & 2.6829878   & 1.4858943 \\
  &  &    &   &     (0.2 \% )  &  (  -0.43 \% ) \\
 & 2.0 & 5.2429941    & 1.4949257  & 5.2418734   & 1.4954987 \\
  &  &    &   &     (0.02 \% )  &  (  -0.04  \% ) \\
 & 5.0 & 13.0092503    & 1.4992028  & 13.0091660   & 1.4992194 \\
  &  &    &   & (0.0006 \% )  &  (  -0.001 \% ) \\
 & 8.0 & 20.7964105     & 1.4996895  & 20.7963895    & 1.4996921 \\
  &  &    &   & (0.0001 \% )  &  ( -0.0002 \% ) \\
3& 0.5 & 1.6413824     & 2.2423840  & 1.6616950    & 2.3943757  \\
 &  &    &   &    (-1.24 \% )  &  (  -6.78 \% ) \\
 & 1.0 & 2.8430931     & 2.3914352  & 2.8296345   & 2.4309959 \\
  &  &    &   &     (0.47 \% )  &  (  -1.65 \% ) \\
 & 2.0 & 5.3400891    & 2.4659363  & 5.3345383   & 2.4710894 \\
  &  &    &   &     (0.10 \% )  &  (   -0.21 \% ) \\
 & 5.0 & 13.0513306   & 2.4941174  & 13.0507979   & 2.4942946 \\
  &  &    &   & (0.004 \% )  &  ( -0.007  \% ) \\
 & 8.0 & 20.8229704    & 2.4976812  & 20.8228335   & 2.4977092 \\
  &  &    &   & (0.0007 \% )  &  (   -0.001 \% ) \\
\end{tabular}
\end{ruledtabular}
\end{table}
\begin{table}
\caption{\label{tab:table2} A sample of Regge pole values for the
electromagnetic field ($2M=1$): Exact versus WKB results.}
\begin{ruledtabular}
\begin{tabular}{cccccc}
&       &\quad Exact  \quad& \quad
Exact \quad&  WKB  &  WKB  \\
 $n $ & $\omega$ &$   \mathrm{Re} \, \lambda_n(\omega)  $
 &  $  \mathrm{Im} \, \lambda_n(\omega)   $
 & $  \mathrm{Re} \, \lambda_n(\omega)    $ &  $ \mathrm{Im} \, \lambda_n(\omega)   $  \\
\hline
1& 0.5 & 1.4815049      & 0.4253267  & 1.4883412    & 0.4258078  \\
 &  &    &   & (-0.46 \% )  &  (    -0.11 \% ) \\
 & 1.0 & 2.7053584     & 0.4754545  & 2.7068558    & 0.4752165 \\
  &  &    &   & (-0.06 \% )  &  (    0.05 \% ) \\
 & 2.0 & 5.2528681     & 0.4932626  & 5.2531027    & 0.4932356 \\
  &  &    &   & (-0.004 \% )  &  (   0.005 \% ) \\
 & 5.0 & 13.0134669     & 0.49889000  & 13.0134830    & 0.49888916 \\
  &  &    &   & (-0.0001 \% )  &  (   0.0002 \% ) \\
 & 8.0 & 20.7990685     & 0.4995649  & 20.7990725    & 0.4995647 \\
  &  &    &   & (-0.00002 \% )  &  (    0.00004  \% ) \\
2& 0.5 & 1.5332591      & 1.3156541  & 1.5406267     & 1.3428745 \\
  &  &    &   & (-0.48 \% )  &  (    -2.07 \% ) \\
 & 1.0 & 2.7692625    & 1.4191713  & 2.7696511    & 1.4211866 \\
  &  &    &   & (-0.01 \% )  &  (  -0.14 \% ) \\
 & 2.0 & 5.2986954     & 1.4736965  & 5.2989080    & 1.4736729 \\
  &  &    &   & (-0.004 \% )  &  (   0.002 \% ) \\
 & 5.0 & 13.0342937     & 1.4953538  & 13.0343184    & 1.4953500 \\
  &  &    &   & (-0.0002 \% )  &  (   0.0003 \% ) \\
 & 8.0 & 20.8122949      & 1.4981621  & 20.8123014     & 1.4981614 \\
  &  &    &   & (-0.00003 \% )  &  (    0.00005 \% ) \\
3& 0.5 & 1.6777130       & 2.1559691  & 1.7082310      & 2.2924069  \\
 &  &    &   & (-1.82 \% )  &  (     -6.33 \% ) \\
 & 1.0 & 2.8922912       & 2.3300925  & 2.8852566     & 2.3612376 \\
  &  &    &   & (0.24 \% )  &  (     -1.34 \% ) \\
 & 2.0 & 5.3854965     & 2.4370870  & 5.3824034    & 2.4403495 \\
  &  &    &   & (0.06 \% )  &  (     -0.13  \% ) \\
 & 5.0 & 13.0753825    & 2.4879569  & 13.0751198    & 2.4880481 \\
  &  &    &   & (0.002 \% )  &  (    -0.004  \% ) \\
 & 8.0 & 20.8385991       & 2.4951766  & 20.8385333  & 2.4951904 \\
  &  &    &   & (0.0003 \% )  &  (     -0.006 \% ) \\
\end{tabular}
\end{ruledtabular}
\end{table}
\begin{table}
\caption{\label{tab:table3} A sample of Regge pole values for
gravitational perturbations ($2M=1$): Exact versus WKB results.}
\begin{ruledtabular}
\begin{tabular}{cccccc}
&       &\quad Exact  \quad& \quad
Exact \quad&  WKB  &  WKB  \\
 $n $ & $\omega$ &$   \mathrm{Re} \, \lambda_n(\omega)  $
 &  $  \mathrm{Im} \, \lambda_n(\omega)   $
 & $  \mathrm{Re} \, \lambda_n(\omega)    $ &  $ \mathrm{Im} \, \lambda_n(\omega)   $  \\
\hline
1& 0.5 & 1.9892654      & 0.3065938  & 2.0250887     & 0.3116324  \\
 &  &    &   & (-1.80 \% )  &  ( -1.64 \%) \\
 & 1.0 & 3.0310943     & 0.4112060  & 3.0434848    & 0.4109477 \\
  &  &    &   & (-0.41 \% )  &  (  0.06  \% ) \\
 & 2.0 & 5.4358037     & 0.4714533  & 5.4381254    & 0.4713563 \\
  &  &    &   & (-0.04 \% )  &  ( 0.02 \% ) \\
 & 5.0 & 13.0897823     & 0.4950209  & 13.0899530    & 0.4950172 \\
  &  &    &   & (-0.001 \% )  &  (   0.0007  \% ) \\
 & 8.0 & 20.8470170     & 0.4980343  & 20.8470594    & 0.4980337 \\
  &  &    &   & (-0.0002 \% )  &  (   0.0001  \% ) \\
2& 0.5 & 1.8354561       & 0.9996004  & 1.9133037      & 1.1052746 \\
  &  &    &   & (-4.24 \% )  &  (   -10.57 \% ) \\
 & 1.0 & 3.0182689     & 1.2608708  & 3.0358421    & 1.2696433 \\
  &  &    &   & (-0.58 \% )  &  ( -0.70 \% ) \\
 & 2.0 & 5.4634287     & 1.4143591  & 5.4679208    & 1.4135833 \\
  &  &    &   & (-0.08 \% )  &  ( 0.05 \% ) \\
 & 5.0 & 13.1090753      & 1.4839779  & 13.1095067     & 1.4839195 \\
  &  &    &   & (-0.003 \% )  &  (  0.004 \% ) \\
 & 8.0 & 20.8598549      & 1.4936074  & 20.8599664     & 1.4935974 \\
  &  &    &   & (-0.0005 \% )  &  (  0.0007 \% ) \\
3& 0.5 & 1.806641        & 1.888817  & 1.929425   & 2.053332  \\
 &  &    &   & (-6.80 \% )  &  (   -8.71 \% ) \\
 & 1.0 & 3.0504514      & 2.1517280  & 3.0724443    & 2.1842791 \\
  &  &    &   & (-0.72 \% )  &  (  -1.51 \% ) \\
 & 2.0 & 5.5221237      & 2.3542390  & 5.5264762    & 2.3546842 \\
  &  &    &   & (-0.08 \% )  &  ( -0.02 \% ) \\
 & 5.0 & 13.1472705    & 2.4697252  & 13.1478689    & 2.4695873 \\
  &  &    &   & (-0.005 \% )  &  (  0.006 \% ) \\
 & 8.0 & 20.8854004      & 2.4877063  & 20.8855672     & 2.4876795 \\
  &  &    &   & (-0.0008 \% )  &  (   0.001 \% ) \\
\end{tabular}
\end{ruledtabular}
\end{table}

\begin{figure}
\includegraphics[height=6cm,width=8cm]{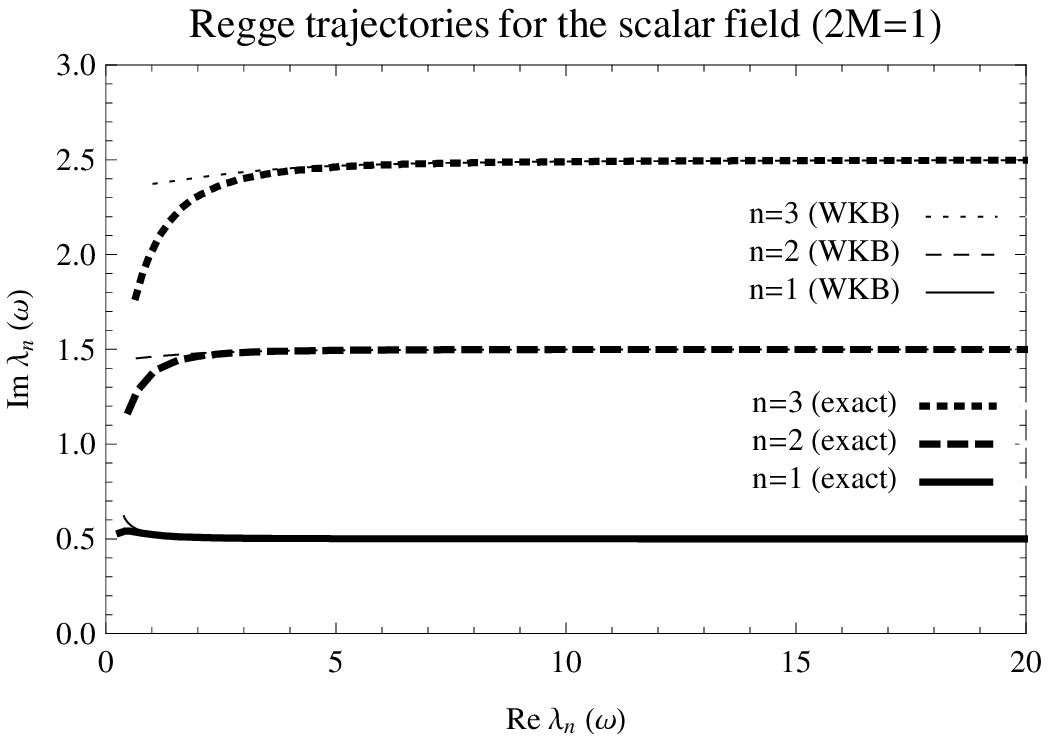}
\includegraphics[height=6cm,width=8cm]{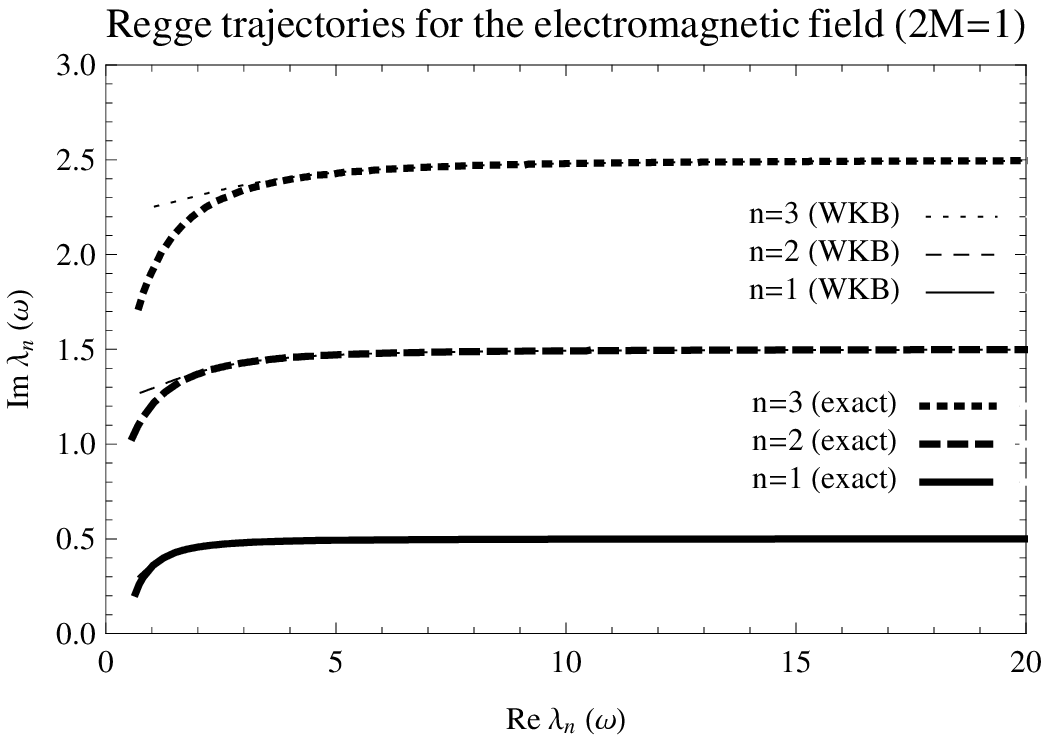}
\includegraphics[height=6cm,width=8cm]{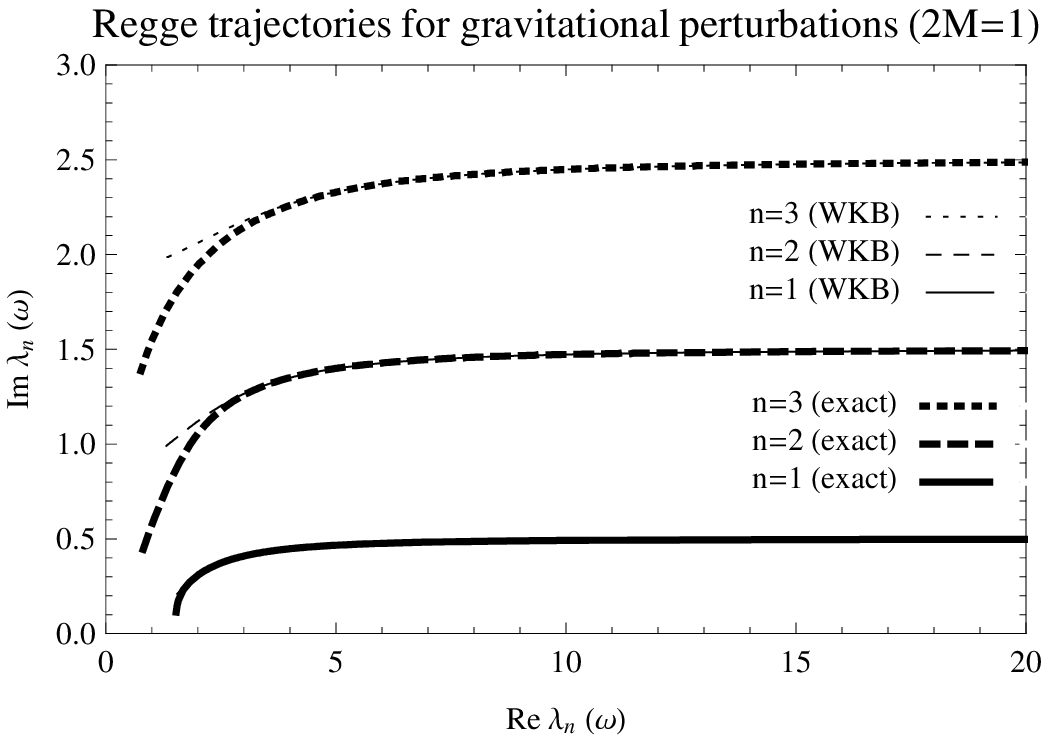}
\caption{\label{fig:TRspin012_EXetWKB} Regge trajectories spins 0, 1
and 2: Exact versus WKB calculations. For the Regge poles
$\lambda_n(\omega)$ ($n=1,2,3$) followed for $\omega=0\to 8$
($2M=1$), $\mathrm{Im} \, \lambda_n(\omega)$ is expressed as a
function of $\mathrm{Re} \, \lambda_n(\omega)$.}
\end{figure}

Tables~\ref{tab:table1}-\ref{tab:table3} present a sample of Regge
pole positions calculated from the WKB result (\ref{PRapproxWKB}).
The WKB results are better for high frequencies and for the first
Regge pole. However, a comparison between the ``exact" and the WKB
results shows a good agreement even for low frequencies and for the
second and third Regge poles. Formula (\ref{PRapproxWKB}) permit us
also to describe very accurately the Regge trajectories of the
Schwarzschild black hole [see Fig.~\ref{fig:TRspin012_EXetWKB}
where, for clarity, we have considered the Regge trajectories in the
form $\mathrm{Im} \, \lambda_n(\omega)$ as a function of
$\mathrm{Re} \, \lambda_n(\omega)$]. It should be also noted that,
in Ref.~\cite{GlampedakisAndersson2003}, Glampedakis and Andersson
have developed a ``quick and dirty" method which provides very
accurate results for the Regge poles (see Table 4 in their article
where the first three Regge poles associated with the scalar field
are given). Their numerical method is certainly better than the WKB
method we have considered. But it is only a purely numerical method
used in order to approximate the exact results. Our WKB approach
permits us to provide a simple and accurate analytical expression
for the Regge poles. It would be also possible to improve
considerably our WKB results by adapting the higher order WKB
approach developed by Konoplya in Ref.~\cite{Konoplya2003}. But we
believe that the Konoplya approach could not provide a formula as
simple as (\ref{PRapproxWKB}).

\begin{figure}
\includegraphics[height=6cm,width=8cm]{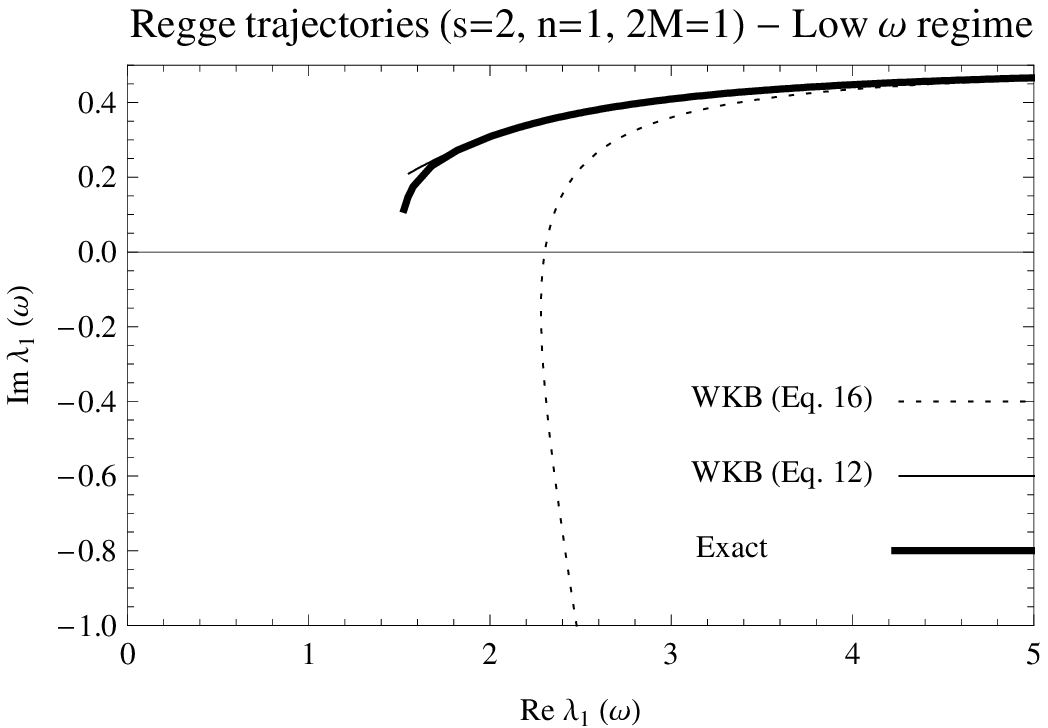}
\caption{\label{fig:TR_s2_n1_ex_et_wkb_etwkbdev} Regge trajectories
(spin 2, $n=1$): Exact versus WKB calculations. The Regge pole
$\lambda_1(\omega)$ followed for $\omega=0.12 \to 1.83$ ($2M=1$).}
\end{figure}

It is moreover possible to simplify (\ref{PRapproxWKB}) and to write
\begin{eqnarray}\label{PRapproxWKB_hf}
&&\lambda_n(\omega) = \left[3\sqrt{3} \, M\omega + \frac{\sqrt{3} \,
a_n}{18 M\omega} \right] \nonumber \\
& & \qquad + i\left(n-1/2 \right) \left[1 + \frac{b_n}{27
(M\omega)^2} \right] + \underset{\omega \to +\infty}{\cal O} \left(
\frac{1}{\omega^3}\right)\nonumber \\
& &
\end{eqnarray}
which is much more elegant. Of course, this alternative formula
provides accurate results only at large $\omega$ and must not be
used at low frequencies (see
Fig.~\ref{fig:TR_s2_n1_ex_et_wkb_etwkbdev}). It provides very simple
expressions for the dispersion relation and the damping of the
``surface waves" lying on the photosphere of the Schwarzschild black
hole. The leading-order terms of Eq.~(\ref{PRapproxWKB_hf})
correspond to the expression (19) of Ref.~\cite{DecaniniFJ_cam_bh}.
Of course, this expression is too coarse to take into account
precisely the spin and index dependence which appear clearly in
Fig.~\ref{fig:PRspins012} and, furthermore, it does not emphasize
the nonlinear behavior of the dispersion relations as well as the
nonconstant behavior of the damping (see
Figs.~\ref{fig:TRspin0}-\ref{fig:TRspin2}). As a consequence, the
numerical calculations which use it are not accurate for low
frequencies $\omega$, for high indices $n$ and for high spins $s$.
For example, for $2M=1$, $\omega =1$, $n=3$, and $s=0$, the formula
(19) of Ref.~\cite{DecaniniFJ_cam_bh} leads to important errors
about $8.6~\%$ for the real part of the Regge pole and $-4.5~\%$ for
its imaginary part. For $s=1$ the errors are, respectively, about
$10.0~\%$ and $-7.3~\%$, and for $s=2$ they are about $14.8~\%$ and
$-16.2~\%$.

In a recent paper \cite{DolanOttewill_2009} written after the first
version of the present one, Dolan and Ottewill have greatly improved
our high frequency approximation (\ref{PRapproxWKB_hf}) by providing
higher orders in $1/\omega$. But, as they have noted, their
expansion is not as accurate as our formula (\ref{PRapproxWKB}) in
the low-$\omega$ regime. This reinforces again our view that
Eq.~(\ref{PRapproxWKB}) is a very interesting result due to its
large range of validity.

In Ref.~\cite{DecaniniFJ_cam_bh}, we have proven that the ``surface
waves" associated with the family of Regge poles propagate on the
photon sphere of the Schwarzschild black hole at $r=3M$. We have
established this connection by using the leading-order behavior of
(\ref{PRapproxWKB_hf}). It is therefore a semiclassical
interpretation which is formally valid for $\omega \to +\infty$. In
this context, we recall (see Ref.~\cite{DecaniniFJ_cam_bh}) that the
resonant part of the form factor of the Schwarzschild black hole is
a superposition of terms like $\exp[i(\lambda_n(\omega)\theta-\omega
t)]$ which can be also written in the form
\begin{equation}\label{prop_sur_photonsphere}
\exp[-k''_n(\omega){\cal L}]\exp[i(k'_n(\omega){\cal L}-\omega t)]
\end{equation}
with ${\cal L}=(3M) \theta$ which denotes the arc length taken on
the photon sphere. Here $k'_n(\omega)=\mathrm{Re} \, \lambda_n
(\omega)/(3M)$ represents the wavenumber of the $n$th ``surface
wave" lying on the photon sphere, or in other terms its dispersion
relation, while $k''_n(\omega)=\mathrm{Im} \, \lambda_n
(\omega)/(3M)$ is its damping constant. To the leading order, the
dispersion relation is linear but if we go beyond this order, it is
clearly nonlinear [see Eqs.~(\ref{PRapproxWKB}) and
(\ref{PRapproxWKB_hf})], and this last result could have important
consequences in strong gravitational lensing. But the previous
analysis could be also inexact: indeed, if we need to take into
account the finite value of the photon frequency, the nonlinear
corrections could also induce a shift for the location of the
unstable circular orbit of the photon and it would be then necessary
to slightly modify (\ref{prop_sur_photonsphere}) and more
particularly the expression of the arc length ${\cal L}$.

Finally, let us note that our WKB results permit us to recover from
the semiclassical formulas (\ref{sc1}) and (\ref{sc4}) well-known
analytical expressions for the QNM complex frequencies. Indeed, by
inserting (\ref{PRapproxWKB}) into Eqs.~(\ref{sc1}) and (\ref{sc4}),
we obtain the well-known large $\ell$ behaviors [see Eq.~(3.1) of
Ref.~\cite{Iyer2}]
\begin{subequations}
\begin{eqnarray}\label{asympQNM}
& & \omega ^{(o)}_{\ell n} = \frac{1}{3\sqrt{3}M} \left[ (\ell +1/2)
-  \frac{a_n}{2 \ell} +  \frac{a_n}{4 \ell^2} + \underset{\ell \to
+\infty}{\cal O}\left( \frac{1}{\ell^3} \right)
 \right]   \label{asympQNM_a}\nonumber \\
& &    \\
 & & \frac{\Gamma _{\ell n}}{2} =
\frac{n-1/2}{3\sqrt{3}M} \left[1+ \frac{c_n}{2 \ell^2} +
\underset{\ell \to +\infty}{\cal O}\left( \frac{1}{\ell^3}
\right)\right] \label{asympQNM_b}
\end{eqnarray}
\end{subequations}
with $\ell \in {\bf N}$ and $n=1, 2, \dots$. Here,
\begin{equation}\label{c_n}
c_n=a_n+2 b_n=-\frac{2}{9}\left[-\frac{235}{432}(n-1/2)^2 +
\frac{1415}{1728} -1 +s^2 \right].
\end{equation}
Here, it is worth recalling that the leading order of
(\ref{asympQNM_a}) can be expressed in terms of the angular velocity
at the unstable circular null geodesic, i.e. at $r=3M$, while the
leading order of (\ref{asympQNM_b}) is related to the corresponding
Lyapunov exponent (see Ref.~\cite{CardosoMirandaBertietal2009}).

\section{Conclusion}

The success of the CAM method in physics is due to its ability to
provide a clear description of a given scattering problem by
extracting the physical information (linked to the geometrical and
diffractive aspects of the scattering process) which is hidden in
partial-wave representations. In the context of black hole physics,
the CAM method has permitted us to clearly interpret the existence
of the weakly damped QNM complex frequencies of the Schwarzschild
black hole in terms of ``surface waves" propagating on/close to its
photon sphere \cite{DecaniniFJ_cam_bh,Andersson1,Andersson2} and, in
the present paper, we have derived analytical expressions for the
dispersion relation and the damping of these ``surface waves" [see
Eqs.~(\ref{PRapproxWKB}) and (\ref{PRapproxWKB_hf})] or,
equivalently, for the location in the CAM plane of the Regge poles
of the Schwarzschild black hole associated with these ``surface
waves".

The analytical expressions we have obtained for the Regge poles of
the Schwarzschild black hole may be helpful in order to simplify
considerably (i) the description of wave scattering from this
gravitational background as well as (ii) the description of the
gravitational radiation created in many black hole processes
involving this gravitational background. In the former context, the
approach initiated by Andersson and Thylwe
\cite{Andersson1,Andersson2} is certainly the most promising way. In
the latter one, a formalism based on the Regge pole machinery has
not yet be developed and all the work remains to be done.

Our results could have also very important consequences and
applications in the context of gravitational lensing. Indeed, until
now the gravitational lens effects due to the Schwarzschild black
hole have been only considered in the framework of geometrical
optics where light propagation is described in terms of null
geodesics. In this framework, photon dispersion is completely
irrelevant. But, in the present paper, we have established from wave
theoretical considerations the existence of dispersion relations for
the photons propagating close to the photon sphere at $r=3M$ and it
is well-known that this ``surface plays" a crucial role in the
context of the strong-deflection gravitational lensing (see, for
example, Ref.~\cite{IyerPetters2007} and references therein). As a
consequence, it seems to us very important to reexamine the
theoretical analysis of gravitational lensing in the light of our
new results. But even without doing such a work, we can already
envision some of their possible observational consequences such as
iridescence phenomenons for the images produced by the Schwarzschild
lens, time delay for photons of different frequencies as well as a
hyperfine structure in the system of images (due to the existence of
an infinity of Regge poles).

In the sixties, high-energy physics was mainly based on the Regge
pole approach of scattering processes (see, for example,
Refs.~\cite{Collins77,Gribov69} and references therein). Of course,
after the discovery of quarks and the development of Yang-Mills
theories, it has taken completely different paths. So, it is amazing
to encounter again the Regge pole concept but now in the context of
gravitation physics (see also Ref.~\cite{DecaniniFolacci2009} where
Regge poles seem to play an important role in the context of the
AdS/CFT correspondence). It is then rather natural to wonder if
there could be something very fundamental behind this which, in
particular, could be helpful in order to understand black holes and
gravitation at the quantum level.

\begin{acknowledgments}
It is a pleasure to acknowledge Bruce Jensen for helpful discussions
on black holes some years ago and Bernard Raffaelli for comments on
the present work. A.F. wishes also to thank Jihad Mourad for his
kind invitation to the APC laboratory where this work was completed.
\end{acknowledgments}

\bibliography{RPandTRforSchwarzschild}

\begin{thebibliography}{28}
\expandafter\ifx\csname natexlab\endcsname\relax\def\natexlab#1{#1}\fi
\expandafter\ifx\csname bibnamefont\endcsname\relax
  \def\bibnamefont#1{#1}\fi
\expandafter\ifx\csname bibfnamefont\endcsname\relax
  \def\bibfnamefont#1{#1}\fi
\expandafter\ifx\csname citenamefont\endcsname\relax
  \def\citenamefont#1{#1}\fi
\expandafter\ifx\csname url\endcsname\relax
  \def\url#1{\texttt{#1}}\fi
\expandafter\ifx\csname urlprefix\endcsname\relax\def\urlprefix{URL }\fi
\providecommand{\bibinfo}[2]{#2}
\providecommand{\eprint}[2][]{\url{#2}}

\bibitem[{\citenamefont{Watson}(1918)}]{Watson18}
\bibinfo{author}{\bibfnamefont{G.~N.} \bibnamefont{Watson}},
  \bibinfo{journal}{Proc.\ R.\ Soc.\ London A} \textbf{\bibinfo{volume}{95}},
  \bibinfo{pages}{83} (\bibinfo{year}{1918}).

\bibitem[{\citenamefont{Sommerfeld}(1949)}]{Sommerfeld49}
\bibinfo{author}{\bibfnamefont{A.}~\bibnamefont{Sommerfeld}},
  \emph{\bibinfo{title}{Partial Differential Equations of Physics}}
  (\bibinfo{publisher}{Academic Press, New York}, \bibinfo{year}{1949}).

\bibitem[{\citenamefont{Newton}(1982)}]{New82}
\bibinfo{author}{\bibfnamefont{R.~G.} \bibnamefont{Newton}},
  \emph{\bibinfo{title}{Scattering Theory of Waves and Particles}}
  (\bibinfo{publisher}{Springer-Verlag, New-York}, \bibinfo{year}{1982}),
  \bibinfo{edition}{2nd} ed.

\bibitem[{\citenamefont{Nussenzveig}(1992)}]{Nus92}
\bibinfo{author}{\bibfnamefont{H.~M.} \bibnamefont{Nussenzveig}},
  \emph{\bibinfo{title}{Diffraction Effects in Semiclassical Scattering}}
  (\bibinfo{publisher}{Cambridge University Press, Cambridge},
  \bibinfo{year}{1992}).

\bibitem[{\citenamefont{W.~T.~Grandy}(2000)}]{Grandy}
\bibinfo{author}{\bibfnamefont{J.}~\bibnamefont{W.~T.~Grandy}},
  \emph{\bibinfo{title}{Scattering of Waves from Large Spheres}}
  (\bibinfo{publisher}{Cambridge University Press, Cambridge},
  \bibinfo{year}{2000}).

\bibitem[{\citenamefont{Collins}(1977)}]{Collins77}
\bibinfo{author}{\bibfnamefont{P.~D.~B.} \bibnamefont{Collins}},
  \emph{\bibinfo{title}{An Introduction to Regge Theory and High-Energy
  Physics}} (\bibinfo{publisher}{Cambridge University Press, Cambridge},
  \bibinfo{year}{1977}).

\bibitem[{\citenamefont{Gribov}(2003)}]{Gribov69}
\bibinfo{author}{\bibfnamefont{V.~N.} \bibnamefont{Gribov}},
  \emph{\bibinfo{title}{The Theory of Complex Angular Momenta: Gribov Lectures
  on Theoretical Physics}} (\bibinfo{publisher}{Cambridge University Press,
  Cambridge}, \bibinfo{year}{2003}).

\bibitem[{\citenamefont{D\'ecanini et~al.}(2003)\citenamefont{D\'ecanini,
  Folacci, and Jensen}}]{DecaniniFJ_cam_bh}
\bibinfo{author}{\bibfnamefont{Y.}~\bibnamefont{D\'ecanini}},
  \bibinfo{author}{\bibfnamefont{A.}~\bibnamefont{Folacci}}, \bibnamefont{and}
  \bibinfo{author}{\bibfnamefont{B.~P.} \bibnamefont{Jensen}},
  \bibinfo{journal}{Phys.\ Rev.\ D} \textbf{\bibinfo{volume}{67}},
  \bibinfo{pages}{124017} (\bibinfo{year}{2003}).

\bibitem[{\citenamefont{Andersson and Thylwe}(1994)}]{Andersson1}
\bibinfo{author}{\bibfnamefont{N.}~\bibnamefont{Andersson}} \bibnamefont{and}
  \bibinfo{author}{\bibfnamefont{K.-E.} \bibnamefont{Thylwe}},
  \bibinfo{journal}{Class.\ Quantum Grav.} \textbf{\bibinfo{volume}{11}},
  \bibinfo{pages}{2991} (\bibinfo{year}{1994}).

\bibitem[{\citenamefont{Andersson}(1994)}]{Andersson2}
\bibinfo{author}{\bibfnamefont{N.}~\bibnamefont{Andersson}},
  \bibinfo{journal}{Class.\ Quantum Grav.} \textbf{\bibinfo{volume}{11}},
  \bibinfo{pages}{3003} (\bibinfo{year}{1994}).

\bibitem[{\citenamefont{Goebel}(1972)}]{Goebel}
\bibinfo{author}{\bibfnamefont{C.~J.} \bibnamefont{Goebel}},
  \bibinfo{journal}{Astrophys.\ J.} \textbf{\bibinfo{volume}{172}},
  \bibinfo{pages}{L95} (\bibinfo{year}{1972}).

\bibitem[{\citenamefont{Ferrari and Mashhoon}(1984)}]{FerrariMashhoon84}
\bibinfo{author}{\bibfnamefont{V.}~\bibnamefont{Ferrari}} \bibnamefont{and}
  \bibinfo{author}{\bibfnamefont{B.}~\bibnamefont{Mashhoon}},
  \bibinfo{journal}{Phys.\ Rev.\ D} \textbf{\bibinfo{volume}{30}},
  \bibinfo{pages}{295} (\bibinfo{year}{1984}).

\bibitem[{\citenamefont{Mashhoon}(1985)}]{Mashhoon85}
\bibinfo{author}{\bibfnamefont{B.}~\bibnamefont{Mashhoon}},
  \bibinfo{journal}{Phys.\ Rev.\ D} \textbf{\bibinfo{volume}{31}},
  \bibinfo{pages}{290} (\bibinfo{year}{1985}).

\bibitem[{\citenamefont{Stewart}(1989)}]{Stewart89}
\bibinfo{author}{\bibfnamefont{J.~M.} \bibnamefont{Stewart}},
  \bibinfo{journal}{Proc.\ R.\ Soc.\ Lond.\ A} \textbf{\bibinfo{volume}{424}},
  \bibinfo{pages}{239} (\bibinfo{year}{1989}).

\bibitem[{\citenamefont{Andersson and Onozawa}(1996)}]{AnderssonOnozawa96}
\bibinfo{author}{\bibfnamefont{N.}~\bibnamefont{Andersson}} \bibnamefont{and}
  \bibinfo{author}{\bibfnamefont{H.}~\bibnamefont{Onozawa}},
  \bibinfo{journal}{Phys.\ Rev.\ D} \textbf{\bibinfo{volume}{54}},
  \bibinfo{pages}{7470} (\bibinfo{year}{1996}).

\bibitem[{\citenamefont{Zerbini and Vanzo}(2004)}]{ZerbiniVanzo2004}
\bibinfo{author}{\bibfnamefont{S.}~\bibnamefont{Zerbini}} \bibnamefont{and}
  \bibinfo{author}{\bibfnamefont{L.}~\bibnamefont{Vanzo}},
  \bibinfo{journal}{Phys.\ Rev.\ D} \textbf{\bibinfo{volume}{70}},
  \bibinfo{pages}{044030} (\bibinfo{year}{2004}).

\bibitem[{\citenamefont{Cardoso et~al.}(2009)\citenamefont{Cardoso, Miranda,
  Berti, Witek, and Zanchin}}]{CardosoMirandaBertietal2009}
\bibinfo{author}{\bibfnamefont{V.}~\bibnamefont{Cardoso}},
  \bibinfo{author}{\bibfnamefont{A.~S.} \bibnamefont{Miranda}},
  \bibinfo{author}{\bibfnamefont{E.}~\bibnamefont{Berti}},
  \bibinfo{author}{\bibfnamefont{H.}~\bibnamefont{Witek}}, \bibnamefont{and}
  \bibinfo{author}{\bibfnamefont{V.~T.} \bibnamefont{Zanchin}},
  \bibinfo{journal}{Phys.\ Rev.\ D} \textbf{\bibinfo{volume}{79}},
  \bibinfo{pages}{064016} (\bibinfo{year}{2009}).

\bibitem[{\citenamefont{Schutz and Will}(1985)}]{SchutzWill}
\bibinfo{author}{\bibfnamefont{B.~F.} \bibnamefont{Schutz}} \bibnamefont{and}
  \bibinfo{author}{\bibfnamefont{C.~M.} \bibnamefont{Will}},
  \bibinfo{journal}{Astrophys.\ J.} \textbf{\bibinfo{volume}{291}},
  \bibinfo{pages}{L33} (\bibinfo{year}{1985}).

\bibitem[{\citenamefont{Iyer and Will}(1987)}]{Iyer1}
\bibinfo{author}{\bibfnamefont{S.}~\bibnamefont{Iyer}} \bibnamefont{and}
  \bibinfo{author}{\bibfnamefont{C.~M.} \bibnamefont{Will}},
  \bibinfo{journal}{Phys.\ Rev.\ D} \textbf{\bibinfo{volume}{35}},
  \bibinfo{pages}{3621} (\bibinfo{year}{1987}).

\bibitem[{\citenamefont{Iyer}(1987)}]{Iyer2}
\bibinfo{author}{\bibfnamefont{S.}~\bibnamefont{Iyer}},
  \bibinfo{journal}{Phys.\ Rev.\ D} \textbf{\bibinfo{volume}{35}},
  \bibinfo{pages}{3632} (\bibinfo{year}{1987}).

\bibitem[{\citenamefont{Bender and Orszag}(1978)}]{BenderOrszag1978}
\bibinfo{author}{\bibfnamefont{C.~M.} \bibnamefont{Bender}} \bibnamefont{and}
  \bibinfo{author}{\bibfnamefont{S.~A.} \bibnamefont{Orszag}},
  \emph{\bibinfo{title}{Advanced Mathematical Methods for Scientists and
  Engineers}} (\bibinfo{publisher}{McGraw-Hill, New-York},
  \bibinfo{year}{1978}).

\bibitem[{\citenamefont{Leaver}(1985)}]{LeaverI}
\bibinfo{author}{\bibfnamefont{E.~W.} \bibnamefont{Leaver}},
  \bibinfo{journal}{Proc.\ R.\ Soc.\ London A} \textbf{\bibinfo{volume}{402}},
  \bibinfo{pages}{285} (\bibinfo{year}{1985}).

\bibitem[{\citenamefont{Majumdar and Panchapakesan}(1989)}]{mp}
\bibinfo{author}{\bibfnamefont{B.}~\bibnamefont{Majumdar}} \bibnamefont{and}
  \bibinfo{author}{\bibfnamefont{N.}~\bibnamefont{Panchapakesan}},
  \bibinfo{journal}{Phys.\ Rev.\ D} \textbf{\bibinfo{volume}{40}},
  \bibinfo{pages}{2568} (\bibinfo{year}{1989}).

\bibitem[{\citenamefont{Glampedakis and
  Andersson}(2003)}]{GlampedakisAndersson2003}
\bibinfo{author}{\bibfnamefont{K.}~\bibnamefont{Glampedakis}} \bibnamefont{and}
  \bibinfo{author}{\bibfnamefont{N.}~\bibnamefont{Andersson}},
  \bibinfo{journal}{Class.\ Quantum Grav.} \textbf{\bibinfo{volume}{20}},
  \bibinfo{pages}{3441} (\bibinfo{year}{2003}).

\bibitem[{\citenamefont{Konoplya}(2003)}]{Konoplya2003}
\bibinfo{author}{\bibfnamefont{R.~A.} \bibnamefont{Konoplya}},
  \bibinfo{journal}{Phys.\ Rev.\ D} \textbf{\bibinfo{volume}{68}},
  \bibinfo{pages}{024018} (\bibinfo{year}{2003}).

\bibitem[{\citenamefont{Dolan and Ottewill}(2009)}]{DolanOttewill_2009}
\bibinfo{author}{\bibfnamefont{S.}~\bibnamefont{Dolan}} \bibnamefont{and}
  \bibinfo{author}{\bibfnamefont{A.~C.} \bibnamefont{Ottewill}},
  \bibinfo{journal}{Class.\ Quantum Grav.} \textbf{\bibinfo{volume}{26}},
  \bibinfo{pages}{225003} (\bibinfo{year}{2009}).

\bibitem[{\citenamefont{Iyer and Petters}(2007)}]{IyerPetters2007}
\bibinfo{author}{\bibfnamefont{S.~V.} \bibnamefont{Iyer}} \bibnamefont{and}
  \bibinfo{author}{\bibfnamefont{A.~O.} \bibnamefont{Petters}},
  \bibinfo{journal}{Gen.\ Rel.\ Grav.} \textbf{\bibinfo{volume}{39}},
  \bibinfo{pages}{1563} (\bibinfo{year}{2007}).

\bibitem[{\citenamefont{D\'ecanini and Folacci}(2009)}]{DecaniniFolacci2009}
\bibinfo{author}{\bibfnamefont{Y.}~\bibnamefont{D\'ecanini}} \bibnamefont{and}
  \bibinfo{author}{\bibfnamefont{A.}~\bibnamefont{Folacci}},
  \bibinfo{journal}{Phys.\ Rev.\ D} \textbf{\bibinfo{volume}{79}},
  \bibinfo{pages}{044021} (\bibinfo{year}{2009}).

\end{thebibliography}

\end{document}